\documentclass[reprint,aps,prl,twocolumn,superscriptaddress,showpacs]{revtex4-1}
\usepackage[dvips]{graphicx}
\usepackage{color}

\begin{document}
\title{Attractive Tomonaga-Luttinger Liquid in a Quantum Spin Ladder}
\author{M. Jeong}
\email{minki.jeong@gmail.com}
\author{H. Mayaffre}
\author{C. Berthier}
\affiliation{Laboratoire National des Champs Magn\'etique Intenses, LNCMI-CNRS (UPR3228), UJF, UPS and INSA, BP 166, 38042, Grenoble Cedex 9, France}
\author{D. Schmidiger}
\affiliation{Neutron Scattering and Magnetism, Laboratory for Solid State Physics, ETH Zurich, Switzerland}
\author{A. Zheludev}
\affiliation{Neutron Scattering and Magnetism, Laboratory for Solid State Physics, ETH Zurich, Switzerland}
\author{M. Horvati\'c}
\email{mladen.horvatic@lncmi.cnrs.fr}
\affiliation{Laboratoire National des Champs Magn\'etique Intenses, LNCMI-CNRS (UPR3228), UJF, UPS and INSA, BP 166, 38042, Grenoble Cedex 9, France}

\begin{abstract}
We present NMR measurements of a strong-leg spin-1/2 Heisenberg antiferromagnetic ladder compound $\mathrm{(C_7H_{10}N)_2CuBr_4}$ under magnetic fields up to 15 T in the temperature range from $1.2$ K down to 50 mK. From the splitting of NMR lines we determine the phase boundary and the order parameter of the low-temperature (3-dimensional) long-range-ordered phase. In the Tomonaga-Luttinger regime above the ordered phase, NMR relaxation reflects characteristic power-law decay of spin correlation functions as $1/T_1\propto T^{1/2K-1}$, which allows us to determine the interaction parameter $K$ as a function of field. We find that field-dependent $K$ varies within the $1<K<2$ range which signifies attractive interaction between the spinless fermions in the Tomonaga-Luttinger liquid.
\end{abstract}

\maketitle

The concept of Tomonaga-Luttinger liquid (TLL) provides a universal description of gapless quantum systems of interacting particles in one dimension (1D), irrespective of the underlying microscopic Hamiltonian \cite{Tomonaga, Haldane81PRL, Giamarchi}. Within the TLL framework, the low-energy properties are completely characterized by only two parameters, the renormalized velocity of excitations $u$ and the dimensionless interaction parameter $K$. For instance, the exponents of all correlation functions, which follow power laws, have simple expressions in terms of $K$ only. This concept has been successfully applied to a wide variety of systems \cite{Giamarchi, Deshpande10Nat}, e.g. organic conductors \cite{Schwartz98PRB}, carbon nanotubes \cite{Bockrath99Nat, Ishii03Nat, Ihara10EPL}, quantum wires \cite{Auslaender02Sci}, edge states of fractional quantum Hall liquid \cite{Grayson98PRL}, cold atoms \cite{Bloch08RMP}, and antiferromagnetic (AF) quantum spin systems \cite{Dender97, Lake05NatMat, Klanjsek08PRL, Ruegg08PRL}. The predicted original properties, such as power-law correlations or fractionalization of the excitations, have been well demonstrated by experiments \cite{Giamarchi, Deshpande10Nat}. Nevertheless, it remains a very difficult task to relate the universal TLL parameters to the microscopic model \cite{Giamarchi}.

The quantum spin systems appear to be a rather unique exception in that respect: their microscopic interactions are often simple and well-defined such that $u$ and $K$ can actually be calculated and directly compared with experiments, which enables quantitative tests of the TLL theory \cite{Giamarchi, Klanjsek08PRL, Ruegg08PRL, Bouillot11PRB}. Among others, spin-1/2 Heisenberg AF ladder systems, having exchange interactions $J_\mathrm{leg}$ along the legs and $J_\mathrm{rung}$ along the rungs, have proven particularly useful under a magnetic field \cite{Klanjsek08PRL, Ruegg08PRL, Bouillot11PRB}. In zero- or low-field, strong quantum fluctuations in a ladder prohibit any magnetic order but lead to a collective singlet ground-state, often called a spin liquid, that is protected by a finite gap to the lowest triplet excitations. Application of a magnetic field $H$ lowers one triplet level by Zeeman energy and eventually closes the gap for the field larger than a critical one, $H>H_{c1}$. The gap closing is accompanied by a transition into a gapless phase that survives up to a saturation field $H_{c2}$. This gapless phase can be described as a TLL of spinless fermions with the help of direct mapping of the spin ladder Hamiltonian onto a model of interacting spinless fermions \cite{Chitra97PRB, Giamarchi99PRB, Furusaki99PRB}. A remarkable feature arising from this mapping concerns the role of the applied field in controlling the TLL physics: $H$, now acting like a chemical potential, controls the filling of a fermion band and determines, together with $J_\mathrm{leg}$ and $J_\mathrm{rung}$, the interaction ($K$) between the fermions \cite{Giamarchi}. The spin ladder in a strong-rung coupling regime ($J_\mathrm{leg}/J_\mathrm{rung}<1$) have been successfully explored in $\mathrm{(C_5H_{12}N)_2CuBr_4}$ (BPCB) \cite{Klanjsek08PRL, Ruegg08PRL, Thielemann09PRB, Bouillot11PRB} which indeed has offered a unique opportunity to test the TLL theory quantitatively over the whole band fillings \cite{Klanjsek08PRL}.

Meanwhile, the ladders are known to provide an interesting contrast for the associated TLLs according to the coupling regime \cite{Giamarchi, Giamarchi99PRB, Furusaki99PRB}: the theory predicts \emph{attractive} spinless fermions (or attractive spinon excitations), i.e. $K>1$, for the ladder in \emph{strong-leg} coupling regime, while repulsive interactions ($K<1$) characterize the strong-rung regime as in BPCB \cite{Klanjsek08PRL}. The prediction for the attractive interactions, however, could not have been confirmed for more than a decade \cite{Giamarchi, Giamarchi99PRB, Furusaki99PRB}, mainly due to the absence of a suitable strong-leg ladder material with low-enough energy scales accessible by available magnetic fields.

\begin{figure*}
\centering
\includegraphics[width=1\textwidth]{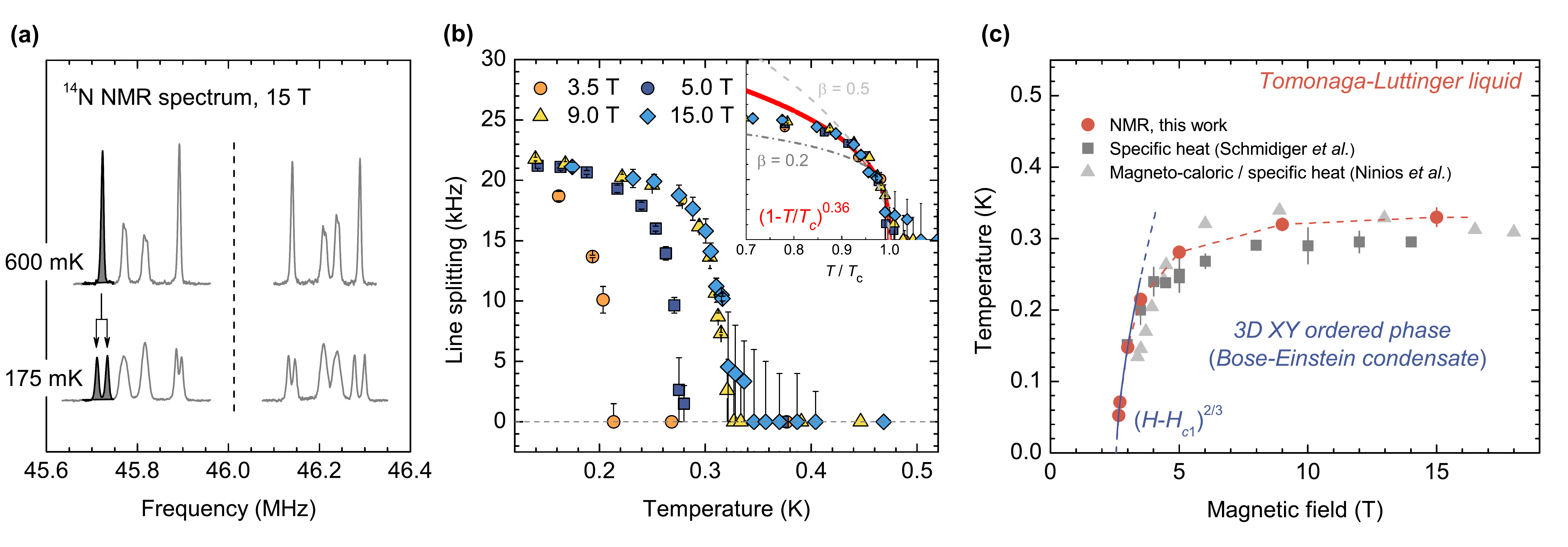}
\caption{(Color online) (a) Typical $^{14}$N NMR spectra above and below the transition, $T_c=340$ mK, in 15 T. Four NMR lines from inequivalent $^{14}$N sites are doubled by quadrupolar splitting ($I=1$), which results in a mirror image about the center (dashed line). The splittings of the lines at low temperature are different due to different hyperfine coupling tensors. (b) The splitting of the lowest-frequency line (filled with grey color in the spectrum shown in (a)) as a function of temperature in four different magnetic fields between 3.5 and 15.0 T. The inset shows fits to the power-law, where solid line represents the best fit. (c) Magnetic phase diagram obtained by plotting the onset temperature ($T_c$) for the line splitting, where the low-field part ($H\leq 3.5$ T) was completed by $^{1}$H NMR (see the text). Solid line represents the critical scaling for magnon BEC. Previous results from thermodynamic measurements, squares for specific heat \cite{Schmidiger12PRL} and triangles for specific heat and magneto-caloric effects \cite{Ninios12PRL}, are also shown.}
\label{Fig1}
\end{figure*}

The recent success \cite{Shapiro08JACS} in synthesis of $\mathrm{(C_7H_{10}N)_2CuBr_4}$, called DIMPY, appears promising in that regard, since this material features a ladder structure of spin-1/2 Cu$^{2+}$ ions coupled via isotropic exchanges, $J_\mathrm{leg}=16.5$ K and $J_\mathrm{rung}=9.5$ K, through Cu$-$Br$-$Br$-$Cu bonds \cite{Shapiro08JACS, Hong10PRL, Schmidiger11PRB, Schmidiger12PRL}. Magnetic one-dimensionality \cite{Hong10PRL} and a small magnon gap of 4.2 K \cite{Hong10PRL, Schmidiger11PRB} were confirmed by neutron scattering. No magnetic transition was observed down to 150 mK in zero field while the field-induced transition into the TLL phase at $H_{c1}=3.0(3)$ T was revealed by specific heat measurements \cite{Hong10PRL}. The density matrix renormalization group (DMRG) calculations, combined with the thermodynamic measurements and neutron scattering results, predicted $H_{c2}\simeq 29$ T \cite{Schmidiger12PRL}. The saturation was indeed observed around a similar value of 31 T at 1.6 K in preliminary pulsed-field magnetization measurements \cite{Jeong}. Furthermore, field-dependent thermodynamic anomalies in specific-heat and magneto-caloric effects were found and attributed to magnetic transition into a low-temperature long-range ordered phase due to weak inter-ladder coupling \cite{Schmidiger12PRL, Ninios12PRL}. This transition is expected to belong to the 3D XY universality class and the ordered phase (canted XY antiferromagnet) can be described as magnon Bose-Einstein condensate (BEC) \cite{Giamarchi99PRB}, though a direct evidence for the order parameter has not yet been found. The available experimental results, supported by theoretical calculations \cite{Schmidiger12PRL}, highlight DIMPY as an ideal strong-leg ladder compound in which one can hope for probing attractive interactions in the TLL phase \cite{Schmidiger12PRL, Ninios12PRL}.

In this Letter, we present the NMR investigation of DIMPY providing the first direct evidence for a TLL with \emph{attractive} interactions. We first identify the order parameter below the magnetic transition temprature $T_c$ through NMR line splitting, and map out the ordered-phase boundary as a function $H$. Then we evidence power-law spin correlations in the TLL phase defined above $T_c$ via the NMR relaxation rate, $1/T_1$, measurements as a function of temperature. They allow us to extract the sign and field-dependent strength of the interaction between the spinless fermions.

For these experiments, we used a single crystal with approximate dimensions $1.5\times 1\times 1$ mm$^3$ and mass 2.5 mg, which has been grown from solution with the temperature gradient method described in detail in Ref. \cite{Yankova12PM}. The monoclinic crystalline structure (space group $P2_1/n$) and quality of the crystal was confirmed by x-ray and neutron scattering \cite{Yankova12PM}, and a trace of paramagnetic impurities were found to be negligible, of the order of 0.1 \% (see the Supplemental Material \cite{SM}). A unit cell contains two different ladders running along the $a$ axis with different rung vectors \cite{Schmidiger11PRB} and each ladder is assigned with two inequivalent N sites (see the Supplemental Material \cite{SM} for the structure). Both $^{14}$N and $^{1}$H NMR were used in complementary manner: simple and well resolved $^{14}$N spectra (Fig. 1a) arising from only four inequivalent N sites in a unit cell allow us to accurately track the line splitting with temperature. Precise temperature-dependence of $1/T_1$ was obtained mainly by $^{1}$H NMR, to take advantage from its strong signal intensity due to the large gyromagnetic ratio  ($^{1}\gamma/^{14}\gamma = 13.8$). We have checked that $^{14}$N and $^{1}$H results are consistent in both spectrum and $T_1$ data as described later.

\begin{figure*}
\centering
\includegraphics[width=1\textwidth]{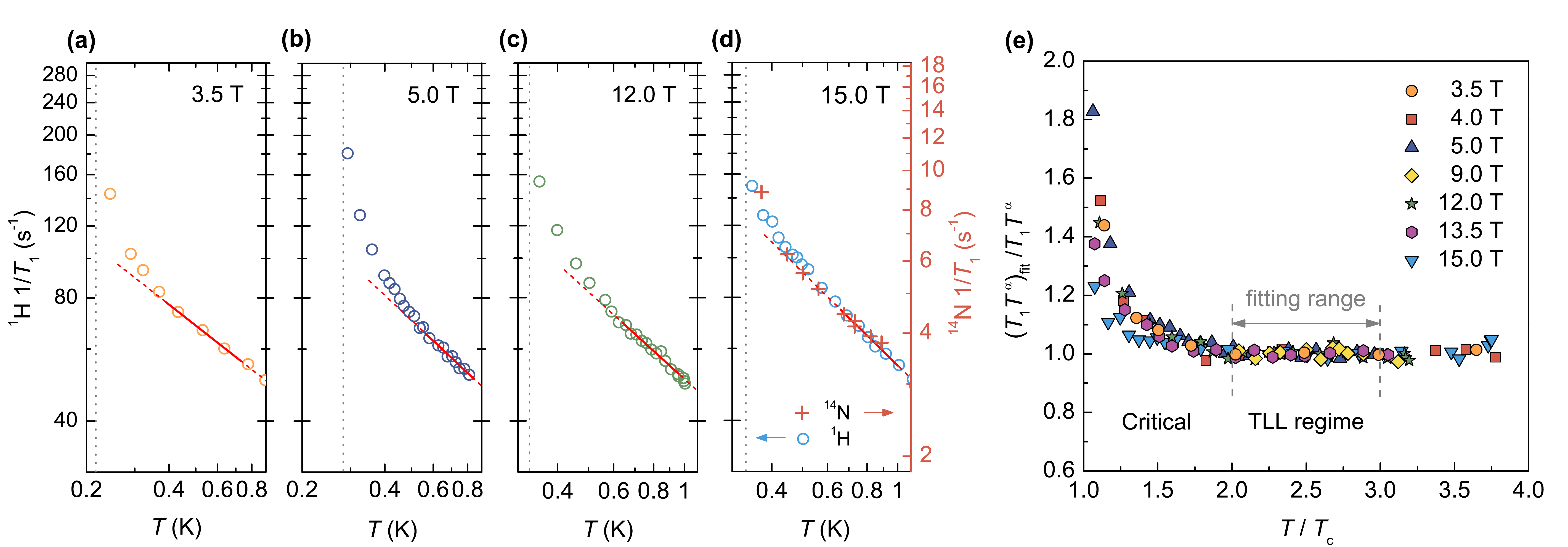}
\caption{(Color online) (a)-(d) $^{1}$H $1/T_1$ (open circles) as a function of temperature at 3.5, 5.0, 12.0, and 15.0 T, respectively. Solid lines are fits to the power-law behavior, $1/T_1\propto T^\alpha$, and represent the fitting range (see the text). Vertical, dotted lines represent $T_c$. The $^{14}$N $1/T_1$ data (crosses) are overlaid in (d) for comparison. (e) Scaled plots of normalized $1/T_1T^\alpha$ as a function of $T/T_c$.}
\label{Fig2}
\end{figure*}

Figure \ref{Fig1}a shows typical $^{14}$N NMR spectra above and below the transition at $T_c=340$ mK in 15 T. The magnetic field was applied along the direction 14$^\mathrm{o}$ off from $a$ axis of the crystal. This choice of orientation ensures that all the $^{14}$N NMR lines are well resolved and separated from one another, so that the line splitting expected from internal fields could be clearly visible. Indeed, in Fig. \ref{Fig1}a each line of the high temperature spectrum splits into two at low temperature. Half of the lines are visibly split, while for the others we simply observed a broadening. The splitting is symmetrical and the split lines have both the shape (Gaussian) and the width identical to the corresponding high-temperature line. This observation points to the development of simple staggered internal fields that should be transverse to the applied field \cite{Giamarchi99PRB}. The line splitting is thus directly proportional to the size of the transverse ordered moments.

We monitored the temperature dependence of the splitting of the lowest-frequency line in the spectrum (Fig. \ref{Fig1}a) in various magnetic fields, which is presented in Fig. \ref{Fig1}b. The data nicely illustrate development of the order parameter upon lowering the temperature. For each field, the onset of the splitting defines $T_c$. The splitting increases rapidly with decreasing temperature and eventually saturates toward 22 kHz around 150 mK. We tried to fit the data to a power-law $\Delta f \propto (1-T/T_c)^\beta$ where $\Delta f$ is the splitting and $\beta$ the critical exponent. As shown in the inset of Fig. \ref{Fig1}b, the best global fitting for the $0.9\leq T/T_c\leq 1$ range gives $\beta = 0.36$ which indeed agrees with the theoretical value of 0.35 for the 3D XY universality class \cite{Campostrini01PRB}. These $^{14}$N NMR spectrum data provide the first direct (microscopic) evidence for the order parameter in DIMPY.

The phase boundary, $T_c(H)$, of the ordered phase is drawn in Fig. \ref{Fig1}c. The low field ($\leq 3.5$ T) part of the phase diagram, where $^{14}$N signal becomes too weak, was completed by monitoring the evolution of $^{1}$H spectrum as a function of temperature or field. The consistency was checked at 3.5 T where both $^{1}$H and $^{14}$N spectra lead to the same $T_c$ value. By extrapolating $T_c(H)$ toward zero, we estimate $H_{c1}=2.55(5)$ T. Solid line represents a critical scaling $T_c \propto (H-H_{c1})^{2/3}$ for a dilute  magnon 3D BEC \cite{Giamarchi99PRB, Nikuni00PRL}. Our $T_c(H)$ data agree with and extend those obtained from the thermodynamic anomalies reported previously \cite{Schmidiger12PRL, Ninios12PRL} (see Fig. \ref{Fig1}c). We thus confirm that those anomalies indeed indicate the magnetic transition into a long-range ordered phase, likely a magnon BEC.

Now we turn our attention to the spin dynamics in the TLL phase above $T_c$. As regards the low-energy excitations, a spin-1/2 ladder can be mapped onto a model of interacting spinless fermions, which, after linearization around the Fermi points and bosonization, transform into the TLL Hamiltonian \cite{Giamarchi}. As remarked before, $K$ in this Hamiltonian measures both the sign and the strength of the interaction and determines the power-law exponents of correlation functions. By NMR $1/T_1$ measurements we probe local spin-spin correlations of the electrons in the low-energy limit. For the spin ladders, in both strong-rung and strong-leg regimes, the transverse correlation functions at $Q=\pi$ are dominant which leads to $1/T_1\propto T^{1/2K-1}$ \cite{Giamarchi99PRB, Hikihara01PRB}. As the field approaches $H_{c1}$ (or $H_{c2}$ as well), the TLL should approach a non-interacting regime where $K=1$ and thus $1/T_1\propto T^{-0.5}$. Independently, a scaling argument for 1D quantum critical regime also leads to $1/T_1\propto T^{-0.5}$ \cite{Orignac07PRB}. Therefore, a continuous variation of $T_1(T)$ is established from the TLL to the 1D quantum critical regime.

Figures \ref{Fig2}a-d show $1/T_1$ of $^{1}$H as a function of temperature obtained in four representative magnetic fields, 3.5, 5.0, 12.0, and 15.0 T, respectively. The $^{14}$N $1/T_1$ data in 15.0 T are also presented in Fig. \ref{Fig2}d to confirm the consistency. The power-law behavior, $1/T_1\propto T^{\alpha}$, is evident for certain temperature ranges above $T_c$. Upon lowering temperature close to $T_c$, thermal critical fluctuations become dominant and enhance $1/T_1$ beyond the TLL behavior. We find the critical fluctuations become negligible above $2\,T_c$, so that we take this value as the lower bound for the validity of the given power-law behavior defining $\alpha(H)$ or $K(H)$. On the high temperature side, the TLL regime is connected to the classical paramagnetic regime through a crossover. We find that power-law behavior is sustained up to above $3\,T_c$ at least. Thus, the power-law exponents $\alpha$ were extracted from the fits to the data systematically in the equivalent temperature range $2\,T_c < T < 3\,T_c$ at all field values.

\begin{figure}
\centering
\includegraphics[width=0.5\textwidth]{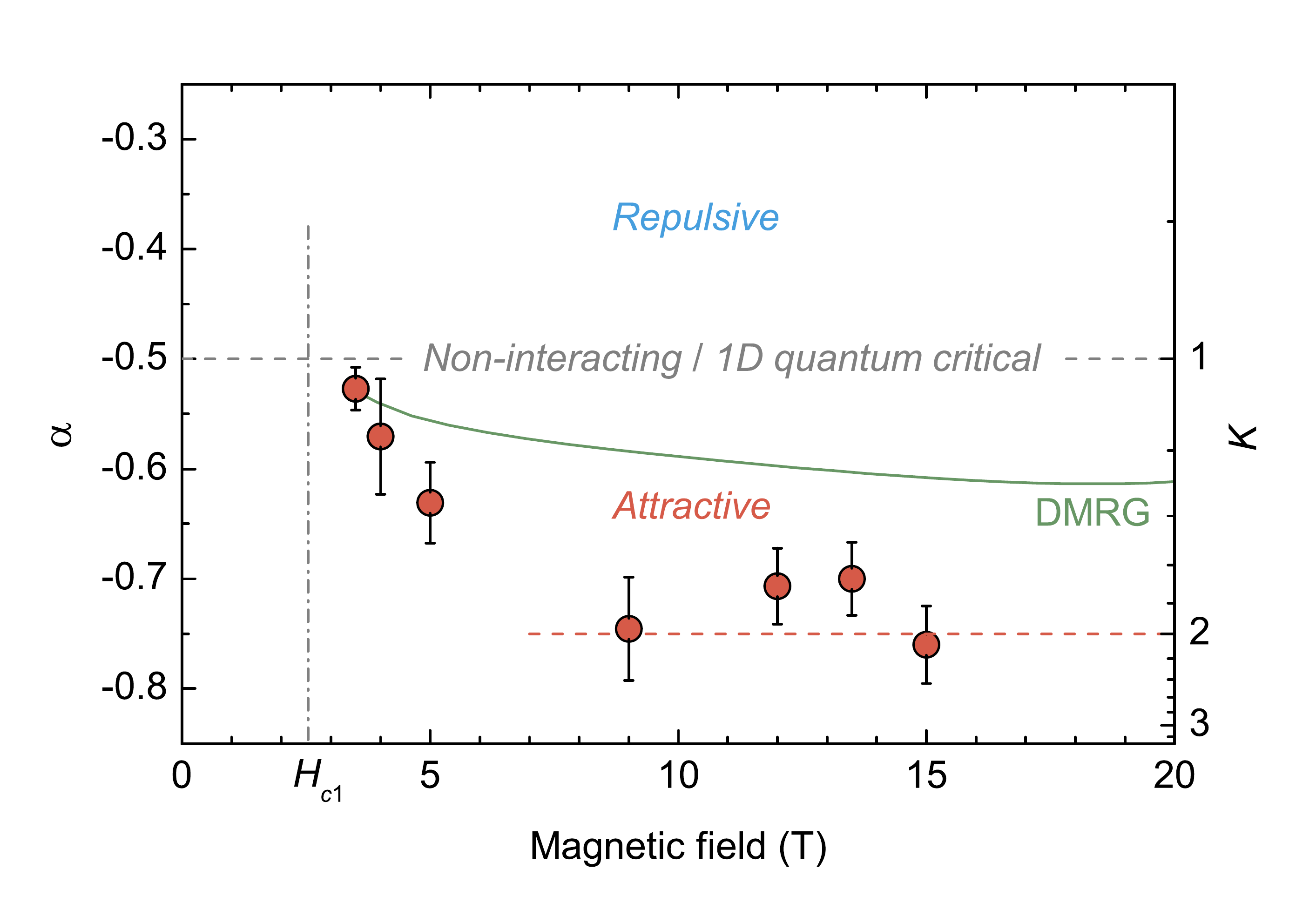}
\caption{(Color online) The extracted $\alpha$ as a function of applied magnetic field. Solid line represents the DMRG calculations reported in Ref. \cite{Schmidiger12PRL}. The corresponding $K$ values can be read from the scale on the right axis}
\label{Fig3}
\end{figure}

To emphasize the universal $1/T_1$ behavior, we present in Fig. \ref{Fig2}e the $1/T_1$ data in a scaled form, that is, by plotting $1/(T_1T^\alpha)$ normalized to 1 within the fitting range as a function of $T/T_c$. We see that the data for $2\,T_c<T<3\,T_c$, and well beyond $3\,T_c$ when available, collapse on a flat line confirming a power-law behavior. The extracted $\alpha$ is plotted as a function of field in Fig. \ref{Fig3}. We find $\alpha$ in the range $-0.75<\alpha<-0.5$, which corresponds to the attractive interaction, i.e. $1<K<2$. Moreover, $\alpha$ clearly converges toward $-0.5$, i.e. $K=1$, as the field approaches $H_{c1}$. This indicates the system approaches the non-interacting regime and/or 1D quantum critical regime, as expected from theory. These NMR relaxation data provide the first direct evidence for power-law correlations in DIMPY and, more generally, attractive interactions in a spin-ladder TLL.

As shown in Fig. \ref{Fig3}, we find a quantitative difference between the experimental results and the DMRG calculations reported in Ref. \cite{Schmidiger12PRL}. The experimental data yield much larger $K$ values than the calculated ones. The reason for this difference is not clear at the moment. In principle, larger $J_\mathrm{leg}/J_\mathrm{rung}$ than estimated could lead to larger $K$. Another possibility might be that the residue of thermal critical fluctuations contribute to $1/T_1$ up to higher temperatures than $2\,T_c$ assumed here, due to their quasi-1D magnetic character. We compared our $1/T_1$ data with the available model of purely 1D ladders with the dominant transverse fluctuations \cite{Giamarchi99PRB, Hikihara01PRB}. However, as one approaches $T_c$ from above, deviation from this simple picture should arise as soon as the inter-ladder correlations become relevant. These correlations can probably be treated in mean-field approximation taking into account the full 1D fluctuations. They are further followed by a crossover regime before reaching the critical one in the vicinity of $T_c$. It will be interesting to take into account these additional theoretical considerations and calculate $1/T_1$ accordingly, to compare with the experimental data as well as to set up the upper bound in temperature for this model. However, these calculations are not available yet and are beyond the scope of the present work. On the experimental side, further measurements using different techniques would help to resolve the issue. Inelastic neutron scattering is another candidate to measure similar properties.

To the best of our knowledge, the results presented here are the first direct evidence for TLL with attractive interactions realized in condensed matter. Most 1D physical realizations such as carbon nanotubes and others are known to support repulsive interactions \cite{Giamarchi}. For AF quantum spin systems, calculations show that a few models, e.g. spin-1/2 XY ladder or XXZ chain with ferromagnetic anisotropy, should support attractive interactions \cite{Giamarchi99PRB, Hikihara01PRB}, but their experimental realization must be quite challenging. Heisenberg spin-1/2 ladders thus appear particularly interesting as they present both repulsive and attractive regimes according to the $J_\mathrm{leg}/J_\mathrm{rung}$ value where the crossover takes place across $J_\mathrm{leg}/J_\mathrm{rung}\simeq 1$. Our $1/T_1$ results now establish this rather unique case of attractive regime for $J_\mathrm{leg}/J_\mathrm{rung}>1$. Furthermore, the magnetic field driven variation of $K$ corresponds to the tuning of TLL model parameters by external means, akin to quantum simulation \cite{Klanjsek12}. Indeed, the spin ladders and other gapped 1D spin systems \cite{Chaboussant98EPJB} increasingly prove their utility under a magnetic field. They host and allow manipulation of quantum phases and many-body phenomena not only at qualitative but also at quantitative level \cite{Klanjsek08PRL, Klanjsek12, Ward13JPCM}, which might be otherwise difficult or impossible to realize.

To summarize, our NMR $1/T_1$ results of the ideal strong-leg spin-$1/2$ ladder compound DIMPY under a magnetic field provide the first evidence for a TLL with attractive interactions. The parameter $K$ is shown to vary between 1 and 2 as a function of field which indicates the field-controlled strength of the attractive interactions. In addition, the NMR line splitting at low temperature evidences the order parameter in DIMPY for the first time. The ordered phase is likely a canted XY antiferromagnet that can be described as the BEC of magnons.

\begin{acknowledgments}
We thank T. Giamarchi for discussions and P. Bouillot for sharing the DMRG data. This work has been supported by EuroMagNET II under the EU contract number 228043 and by the Swiss National Fund under MaNEP and Division II.
\end{acknowledgments}

\newpage
\section*{Supplemental Matierial}
In this Supplemental Material, we provide some useful material properties of Bis(2,3-dimethylpyridinium) Tetrabromocuprate with the chemical formula $\mathrm{(C_7H_{10}N)_2CuBr_4}$ (DIMPY for short) and basic information about our single-crystal sample used in the present study.

\begin{figure}
\centering
\includegraphics[width=1\columnwidth]{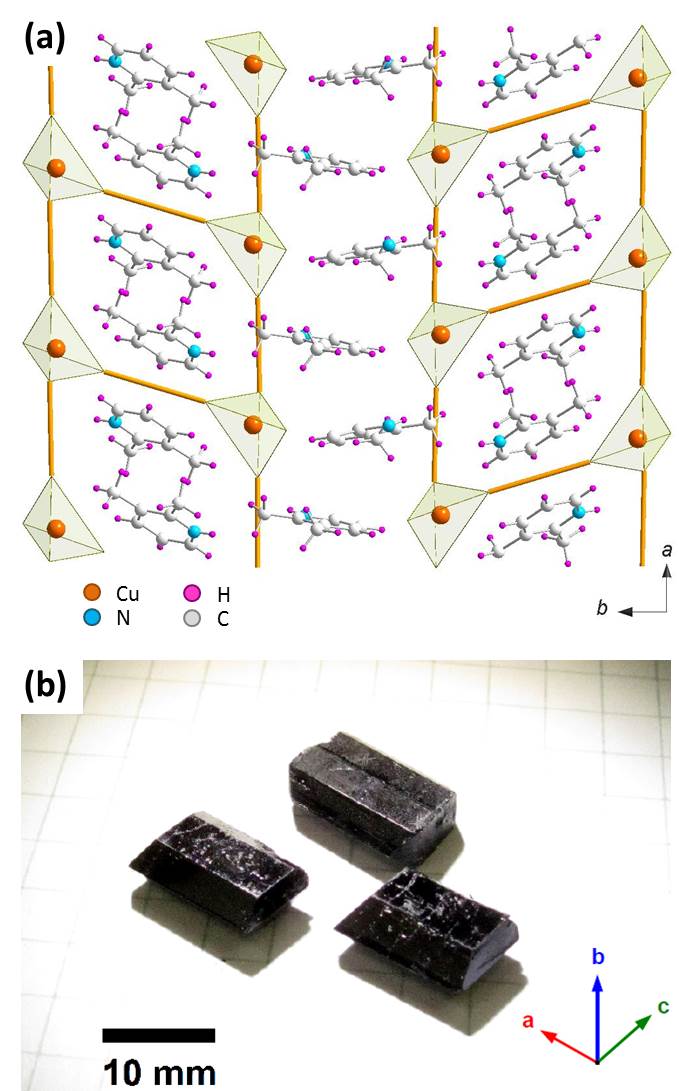}
\caption{(a) Crystallographic structure of (C$_7$H$_{10}$N)$_2$CuBr$_4$, as seen from the crystallographic $c$ axis (b) Large single-phase crystals grown from solution
according to the temperature-gradient method.} \label{crystal}
\end{figure}

DIMPY is an organometallic material based on the organic 2,3-dimethylpyridine molecules and tetrahedral $[\mathrm{CuBr_4}]^{2-}$ units. It crystallizes into a monoclinic structure of which space group is $P2_1/n$ and lattice constants $a=7.504$, $b=31.61$, and $c=8.202$, with $\beta=98.97$ \cite{Shapiro08JACS}. Figure \ref{crystal}a shows the crystal structure. The tetrahedral $[\mathrm{CuBr_4}]^{2-}$ units (Cu ions are represented by orange spheres and Br ions are on the corner) form ladder structures depicted by thick yellow lines. The magnetic behavior is goverened by the $\mathrm{Cu}^{2+}$ ions embedded in a tetrahedral $\mathrm{Br}^{-}$ environment. The $\mathrm{Cu}^{2+}$ ions interact with one another through Cu$-$Br$-$Br$-$Cu superexchange path and build two decoupled ladder-like networks along the crystallographic $a$ axis. These two ladders appear with different rung vectors $\mathbf{d}_{1,2}=(0.423,\pm0.256,0.293)$ in fractional coordinates \cite{Schmidiger11PRB}.

The single crystals were grown in our laboratory at ETH, Z\"urich. Figure \ref{crystal}b shows a photograph of the large single-phase crystals (of mass up to 1.5 g) grown from solution by the temperature gradient method \cite{Yankova12PM}. The unit cell, crystallographic axes, and orientation were checked by x-ray and found in agreement with those reported earlier by others. 
\begin{figure}
\centering
\includegraphics[width=1\columnwidth]{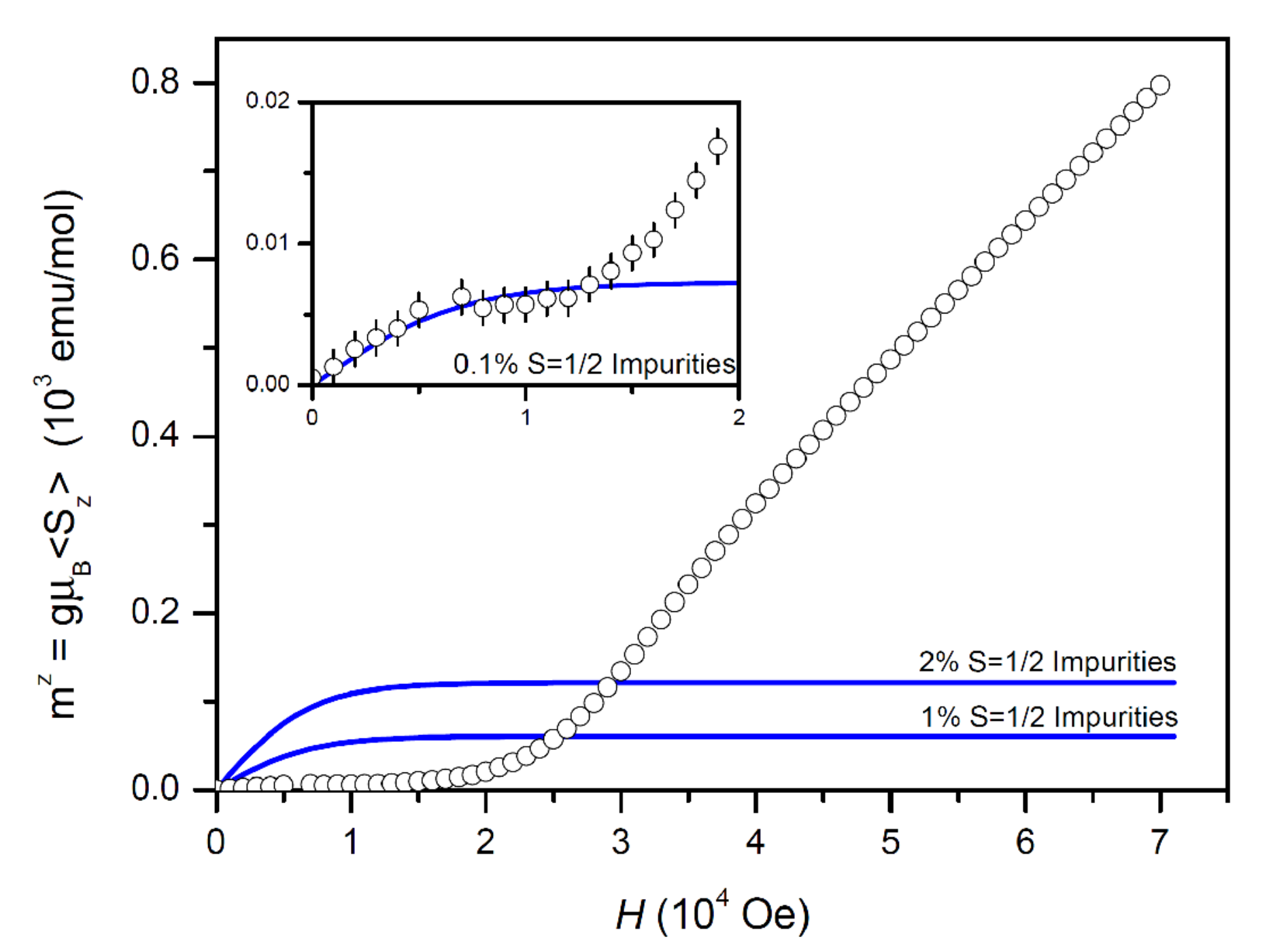}
\caption{Induced magnetization of DIMPY (open symbols) as a function of magnetic field, measured with a MPMS SQUID device ($T=500$~mK and $H||a$). Solid lines correspond to calculated paramagnetic $S=1/2$ contribution for a concentration of 1\% and 2\% as well as the fitted value of 0.1\% (inset).} \label{Purity}
\end{figure}

In order to check the level of paramagnetic impurities, we performed magnetization measurements at $T=500$ mK. A single crystal (7.6 mg) grown from the same solution as that used for the NMR experiments was measured. The sample was oriented such that the field is along the crystallographic $a$-axis. Figure \ref{Purity} shows the obtained magnetization data. Solid lines are the calculations for 1 \% and 2 \% of $S=1/2$ paramagnetic impurities, which are shown for comparison. The inset shows an enlarged view of the low-field part where the data at low fields follow the saturation behavior calculated for 0.1 \% of impurities (solid line). This result shows that in our single crystal the concentration of magnetic impurities is negligible.

\begin{figure}
\centering
\includegraphics[width=1.05\columnwidth]{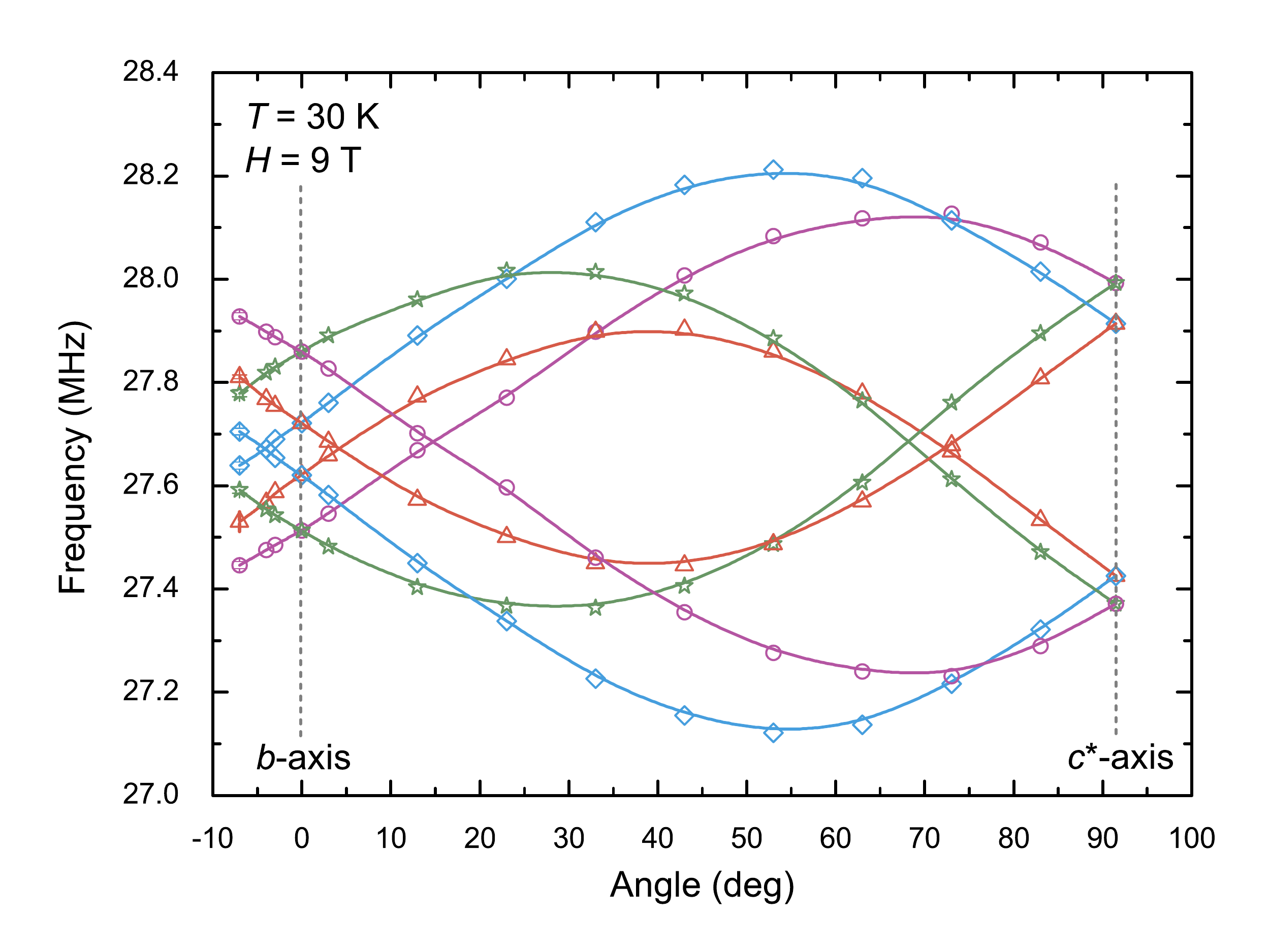}
\caption{Rotation pattern of the $^{14}$ NMR spectrum for the magnetic field directions from the crystallographic $b$ to $c^*$ axis. The data were obtained at 30 K in 9 T. The lines are guides to the eye. The different colors and symbols represent the inequivalent sites.} \label{rotation}
\end{figure}

For the NMR measurements, we used both $^{14}$N and $^{1}$H nuclei which are depicted by pink and light blue spheres in Fig. \ref{crystal}a, respectively. Both these nuclei belong to the ligand molecules. As shown in Fig. \ref{crystal}a, there are four inequivalent N sites in a unit cell. In general, all these sites can be resolved in NMR spectra when a magnetic field is applied off the symmetry axes, which is the configuration used in the present study. Since $I=1$ for $^{14}$N nuclei, quadrupolar splitting doubles the number of NMR lines resulting in a mirror image of the NMR spectrum having total of eight NMR lines. On the other hand, if a field is applied along the symmetry axes, the number of lines reduces to four. These features are clearly shown in Fig. \ref{rotation} presenting the angle dependence of NMR spectra for rotation from $b$ axis to $c^*$ axis obtained at 30 K in a magnetic field of 9 T. Four NMR lines, observed when the crystallographic $b$ axis is along the field, are split into eight as one rotates the sample with respect to the field. The resulting rotation pattern mainly arises from the quadrupolar interactions between the nuclei and the crystal electric-field gradient,  whereas the contributions from magnetic dipolar interactions between the nuclei and the electrons are smaller. The eight lines eventually converges back toward four lines again as the sample's $c$ axis coincides with the field. As explained in the main text, since we were interested in determining the splitting of a single line due to internal fields, we intentionally tilted the sample off the symmetry axes in order to maximize the resolution.


\begin{thebibliography}{99}
\bibitem{Tomonaga}S. Tomonaga, Prog. Theor. Phys. {\bf 5,} 544 (1950); J. M. Luttinger, J. Math. Phys. {\bf 4,} 1154 (1963); D. C. Mattis and E. H. Lieb, J. Math. Phys. {\bf 6,} 304 (1965).
\bibitem{Haldane81PRL}F. D. M. Haldane, Phys. Rev. Lett. {\bf 47,} 1840 (1981).
\bibitem{Giamarchi}T. Giamarchi, {\it Quantum Physics in One Dimension} (Oxford University, Oxford, 2004).
\bibitem{Deshpande10Nat}V. V. Deshpande, M. Bockrath, L. I. Glazman, and A. Yacoby, Nature {\bf 464,} 209 (2010).
\bibitem{Schwartz98PRB} A. Schwartz, M. Dressel, G. Gr\"{u}ner, V. Vescoli, L. Degiorgi, and T. Giamarchi, Phys. Rev. B {\bf 58,} 1261 (1998).
\bibitem{Bockrath99Nat}M. Bockrath, D. H. Cobden, J. Lu, A. G. Rinzler, R. E. Smalley, L. Balents and P. L. McEuen, Nature {\bf 397,} 598 (1999).
\bibitem{Ishii03Nat} H. Ishii, H. Kataura, H. Shiozawa, H. Yoshioka, H. Otsubo, Y. Takayama, T. Miyahara, S. Suzuki, Y. Achiba, M. Nakatake, T. Narimura, M. Higashiguchi, K. Shimada, H. Namatame, and M. Taniguchi, Nature {\bf 426,} 540 (2003).
\bibitem{Ihara10EPL}Y. Ihara, P. Wzietek, H. Alloul, M. H. R\"{u}mmeli, Th. Pichler, and F. Simon, Euro. Phys. Lett. {\bf 90,} 17004 (2010).
\bibitem{Auslaender02Sci} O. M. Auslaender, A. Yacoby, R. de Picciotto, K. W. Baldwin, L. N. Pfeiffer, and K. W. West, Science {\bf 295,} 825 (2002).
\bibitem{Grayson98PRL}M. Grayson, D. C. Tsui, L. N. Pfeiffer, K. W. West, and A. M. Chang, Phys. Rev. Lett. {\bf 80,} 1062 (1998).
\bibitem{Bloch08RMP} I. Bloch, J. Dalibard, and W. Zwerger, Rev. Mod. Phys. {\bf 80,}
885 (2008).
\bibitem{Dender97}D. C. Dender, Ph.D. thesis, Johns Hopkins University, 1997.
\bibitem{Lake05NatMat} B. Lake, D. A. Tennant, C. D. Frost, and S. E. Nagler, Nature Mater. {\bf 4,} 329 (2005).
\bibitem{Klanjsek08PRL}M. Klanj\v{s}ek, H. Mayaffre, C. Berthier, M. Horvati\'c, B. Chiari, O. Piovesana, P. Bouillot, C. Kollath, E. Orignac, R. Citro, and T. Giamarchi, Phys. Rev. Lett. {\bf 101,} 137207 (2008).
\bibitem{Ruegg08PRL}Ch. R\"{u}egg, K. Kiefer, B. Thielemann, D. F. McMorrow, V. Zapf, B. Normand, M. B. Zvonarev, P. Bouillot, C. Kollath, T. Giamarchi, S. Capponi, D. Poilblanc, D. Biner, and K. W. Kr\"{a}mer, Phys. Rev. Lett. {\bf 101,} 247202 (2008).
\bibitem{Bouillot11PRB}P. Bouillot, C. Kollath, A. M. L\"{a}uchli, M. Zvonarev, B. Thielemann, Ch. R\"{u}egg, E. Orignac, R. Citro, M. Klanj\v{s}ek, C. Berthier, M. Horvati\'c, and Thierry Giamarchi, Phys. Rev. B {\bf 83,} 054407 (2011).
\bibitem{Giamarchi99PRB}T. Giamarchi and A.M. Tsvelik, Phys. Rev. B {\bf 59,} 11398 (1999).
\bibitem{Furusaki99PRB}A. Furusaki and S. C. Zhang, Phys. Rev. B {\bf 60,} 1175 (1999).
\bibitem{Chitra97PRB}R. Chitra and T. Giamarchi, Phys. Rev. B {\bf 55,} 5816 (1997).
\bibitem{Thielemann09PRB}B. Thielemann, Ch. R\"{u}egg, K. Kiefer, H. M. R{\o}nnow, B. Normand, P. Bouillot, C. Kollath, E. Orignac, R. Citro, T. Gimamarchi, A. M. L\"{a}uchli, D. Biner, K. W. Kr\"{a}mer, F. Wolff-Fabris, V. S. Zapf, M. Jaime, J. Stahn, N. B. Christensen, B. Grenier, D. F. McMorrow, and J. Mesot, Phys. Rev. B {\bf 79,} 020408(R) (2009).
\bibitem{Shapiro08JACS}A. Shapiro, C. P. Landee, M. M. Turnbull, J. Jornet, M. Deumal, J. J. Novoa, M. A. Robb, and W. Lewis, J. Am. Chem. Soc. {\bf 129,} 952 (2007).
\bibitem{Hong10PRL}T. Hong, Y. H. Kim, C. Hotta, Y. Takano, G. Tremelling, M. M. Turnbull, C. P. Landee, H.-J. Kang, N. B. Christensen, K. Lefmann, K. P. Schmidt, G. S. Uhrig, and C. Broholm, Phys. Rev. Lett. {\bf 105,} 137207 (2010).
\bibitem{Schmidiger11PRB}D. Schmidiger, S. M\"{u}hlbauer, S. N. Gvasaliya, T. Yankova, and A. Zheludev, Phys. Rev. B {\bf 84,} 144421 (2011).
\bibitem{Schmidiger12PRL}D. Schmidiger, P. Bouillot, S. M\"{u}hlbauer, S. Gvasaliya, C. Kollath, T. Giamarchi, and A. Zheludev, Phys. Rev. Lett. {\bf 108,} 167201 (2012).
\bibitem{Jeong}M. Jeong {\it et al}., unpublished.
\bibitem{Ninios12PRL}K. Ninios, T. Hong, T. Manabe, C. Hotta, S. N. Herringer, M. M. Turnbull, C. P. Landee, Y. Takano, and H. B. Chan, Phys. Rev. Lett. {\bf 108,} 097201 (2012). 
\bibitem{Campostrini01PRB}M. Campostrini, M. Hasenbusch, A. Pelissetto, P. Rossi, and E. Vicari, Phys. Rev. B {\bf 63,} 214503 (2001).
\bibitem{Nikuni00PRL}T. Nikuni, M. Oshikawa, A. Oosawa, and H. Tanaka, Phys. Rev. Lett. {\bf 84,} 5868 (2000).
\bibitem{Hikihara01PRB}T. Hikihara and A. Furusaki, Phys. Rev. B {\bf 63,} 134438 (2001).
\bibitem{Orignac07PRB}E. Orignac, R. Citro, and T. Giamarchi, Phys. Rev. B {\bf 75,} 140403(R) (2007).
\bibitem{Yankova12PM}T. Yankova, D. H\"{u}vonen, S. M\"{u}hlbauer, D. Schmidiger, E. Wulf, S. Zhao, A. Zheludev, T. Hong, V. O. Garlea, R. Custelcean and G. Ehlers, Philosophical Magazine {\bf 92,} 2629 (2012).
\bibitem{SM}See the Supplemental Material for experimental details on the sample.
\bibitem{Klanjsek12}M. Klanj\v{s}ek, M. Horvati\'c, C. Berthier, H. Mayaffre,  E. Canevet, B. Grenier, P. Lejay, and E. Orignac, arXiv:1202.6374.
\bibitem{Chaboussant98EPJB}G. Chaboussant, M.-H. Julien, Y. Fagot-Revurat, M. Hanson, L. P. L\'evy, C. Berthier, M. Horvati\'c, and O. Piovesana, Eur. Phys. J. B {\bf 6,} 167 (1998).
\bibitem{Ward13JPCM}S. Ward, P. Bouillot, H. Ryll, K. Kiefer, K. W. Kr\"{a}mer, Ch. R\"{u}egg, C. Kollath, and T. Giamarchi, J. Phys.: Condens. Matter {\bf 25,} 014004 (2013).
\end{thebibliography}

\begin{thebibliography}{99}
\bibitem{Shapiro08JACS}A. Shapiro, C. P. Landee, M. M. Turnbull, J. Jornet, M. Deumal, J. J. Novoa, M. A. Robb, and W. Lewis, J. Am. Chem. Soc. {\bf 129,} 952 (2007).
\bibitem{Schmidiger11PRB}D. Schmidiger, S. M\"{u}hlbauer, S. N. Gvasaliya, T. Yankova, and A. Zheludev, Phys. Rev. B {\bf 84,} 144421 (2011).
\bibitem{Yankova12PM}T. Yankova, D. H\"{u}vonen, S. M\"{u}hlbauer, D. Schmidiger, E. Wulf, S. Zhao, A. Zheludev, T. Hong, V. O. Garlea, R. Custelcean and G. Ehlers, Philosophical Magazine {\bf 92,} 2629 (2012).
\bibitem{Schmidiger12PRL}D. Schmidiger, P. Bouillot, S. M\"{u}hlbauer, S. Gvasaliya, C. Kollath, T. Giamarchi, and A. Zheludev, Phys. Rev. Lett. {\bf 108,} 167201 (2012).
\end{thebibliography}
\end{document}